\documentclass[a4paper,twocolumn]{revtex4}
\usepackage[utf8]{inputenc}
\usepackage[english]{babel}
\usepackage[T1]{fontenc}
\usepackage{amsfonts}
\usepackage[dvips]{graphicx}
\usepackage[usenames]{color}
\usepackage{amssymb,amsmath,amsthm}
\usepackage{hyperref}
\DeclareMathOperator{\Tr}{Tr}
\DeclareMathOperator{\supp}{supp}

\newtheorem{proposition}{Proposition}

\begin{document}
\title{Probability-fidelity tradeoffs for targeted quantum operations}

\author{G. M. D'Ariano} \affiliation{Quit group, Dipartimento di
  Fisica ``A. Volta'', via A. Bassi 6, 27100 Pavia,
  Italy}\affiliation{Istituto Nazionale di Fisica Nucleare, Gruppo IV,
  via A. Bassi 6, 27100 Pavia, Italy}

\author{S. Facchini} \affiliation{Quit group, Dipartimento di
  Fisica ``A. Volta'', via A. Bassi 6, 27100 Pavia,
  Italy}\affiliation{Istituto Nazionale di Fisica Nucleare, Gruppo IV,
  via A. Bassi 6, 27100 Pavia, Italy}

\author{P. Perinotti} \affiliation{Quit group, Dipartimento di
  Fisica ``A. Volta'', via A. Bassi 6, 27100 Pavia,
  Italy}\affiliation{Istituto Nazionale di Fisica Nucleare, Gruppo IV,
  via A. Bassi 6, 27100 Pavia, Italy}

\author{M. F. Sacchi} \affiliation{Quit group, Dipartimento di
  Fisica ``A. Volta'', via A. Bassi 6, 27100 Pavia,
  Italy}\affiliation{CNR-INFM and CNISM, Unit\`a di Pavia, 
  via A. Bassi 6, 27100 Pavia, Italy}

\begin{abstract}
  We present probability-fidelity tradeoffs for a varying quantum
  operation with fixed input-output states and for a varying inversion
  of a fixed quantum operation.
\end{abstract}

\maketitle

\newcommand{\ket}[1]{| #1 \rangle}
\newcommand{\bra}[1]{\langle #1 |}
\newcommand{\Ket}[1]{| #1 \rangle \! \rangle}
\newcommand{\Bra}[1]{\langle \! \langle #1 |}
\newcommand{\BraKet}[2]{\langle \! \langle #1 | #2 \rangle  \! \rangle}
\newcommand{\KetBra}[2]{\Ket{#1} \Bra{#2}}
\newcommand{\braket}[2]{\langle #1 | #2 \rangle}
\newcommand{\ketbra}[2]{\ket{#1} \bra{#2}}
\newcommand{\hilb}[1]{\mathcal{#1}}

\def\N#1{\Vert{#1}\Vert}
\def\refp#1{(\ref{#1})}
\def\>{\rangle} \def\<{\langle}

\section{Introduction}

Since the seminal work in Ref. \cite{fp96}, many
information/disturbance tradeoffs have been derived in a wide range of
frameworks \cite{fu98,ba00,banaszek01.prl, banaszek01.pra,
  mista05.pra,max,cv,paris,bs06,maxun}.  Despite this variety, all
tradeoffs were derived based on figures of merit defined as
average over some ensemble, e.g. the uniform ensemble of all
transformations.

In this paper, following the suggestions contained in
Ref. \cite{da03}, we study the behavior of a \emph{single} quantum
operation in some simple cases, along the following lines. After
reviewing the probability of transforming a pair of pure states to
another given pair \cite{cb98}, we extend it to mixed target states,
and then we provide a tradeoff between the probability and the
fidelity of such a transformation. Finally, we present the
probability-fidelity tradeoff in the inversion of an atomic
(i.~e. single-Kraus) quantum operation.

\section{State transformations}

We are given an ensemble $\mathsf{E}=\{q_\pm,
\ketbra{\psi_\pm}{\psi_\pm}\}$ of two pure states $\ket{\psi_\pm}$
with equal \emph{a priori} probabilities $q_\pm=1/2$, and a pair of
(generally mixed) target states $\rho_\pm$. We want to find a quantum
operation which realizes the transformation
\begin{equation}
\ket{\psi_\pm} \longrightarrow \rho_\pm
\end{equation}
maximizing the mean probability of success over the ensemble.

For pure final states $\rho_\pm = \ketbra{\phi_\pm}{\phi_\pm}$ the
problem has been solved in Ref. \cite{cb98}:
\begin{proposition}\label{thm-pairtrans}
The maximum mean probability is
\begin{equation}\label{eqn-chefles}
  p=\min \left\lbrace \frac{1-\vert\braket{\psi_+}{\psi_-}\vert}{1-\vert\braket{\phi_+}{\phi_-}\vert}, 1 \right\rbrace.
\end{equation}
Moreover, this probability is achieved with a balanced transformation,
i.e. a transformation occuring with equal probability on both initial
states.
\end{proposition}

Indeed, the above proposition can be extended also to final mixed states.
\begin{proposition}\label{thm-pairmixtrans}
For generally mixed final states $\rho_\pm$ the maximum mean probability is
\begin{equation} \label{eq-maxprob}
  p = \min \left\lbrace \frac{1-\vert\braket{\psi_+}{\psi_-}\vert}{1-F(\rho_+, \rho_-)}, 1\right\rbrace
\end{equation}
where $F(\rho,\sigma):=\Tr\sqrt{\sqrt{\rho}\sigma \sqrt{\rho}}$
is the Uhlmann fidelity \cite{uh76}.  Moreover, the probability is
achieved with a balanced transformation.
\begin{proof}
Suppose we have a quantum operation $\mathcal{E}$ realizing the transformation
\begin{equation}
\ket{\psi_\pm} \longrightarrow \rho_\pm,
\end{equation}
with certain probabilities $p_\pm$. Using the Ozawa dilation theorem
\cite{ozawa} for quantum instruments we can realize the quantum
operation in the following way
\begin{equation}
\mathcal{E}(\rho)=\Tr_2 [(I\otimes P)U (\rho\otimes\vert 0\rangle\langle 0|) U^\dag (I\otimes P)]
\end{equation}
where $|0\>$ is any pure state of an anciallary system, $U$ is a
unitary system-ancilla interaction, $P$ is an orthogonal projector,
and we take the trace on the ancilla. Since unitaries and projectors
cannot turn a pure state into a mixed one, the quantum operation
$\mathcal{E}$, when applied to the our initial states
$\ket{\psi_\pm}$, will have the form
\begin{equation}
\mathcal{E}(\ketbra{\psi_\pm}{\psi_\pm}) = p_\pm \Tr_2 (\ketbra{\Phi_\pm}{\Phi_\pm})
\end{equation}
where $\ket{\Phi_\pm}$ are joint ancilla-system states and $p_\pm$ are
the success probabilities.  Note that $\ket{\Phi_\pm}$ are actually
purifications of the final states $\rho_\pm$.

In this way we proved that every state transformation $\ket{\psi_\pm}
\rightarrow \rho_\pm$ can be realized with a transformation between
pure states $\ket{\psi_\pm} \rightarrow \ket{\Phi_\pm}$ followed by a
partial trace. Thus, in order to maximize the probability of
$\ket{\psi_\pm} \rightarrow \rho_\pm$ it is not restrictive to search
only among those transformations which take $\ket{\psi_\pm}$ into
purifications of the final states $\rho_\pm$.

From Uhlmann's theorem \cite{uh76} we have that
\begin{equation}\label{eqn-uhl}
\vert\braket{\Phi_+}{\Phi_-}\vert \leq F(\rho_+, \rho_-),
\end{equation}
for all the purifications of $\rho_\pm$, and thus
\begin{equation}
  \frac{1-\vert\braket{\psi_+}{\psi_-}\vert}{1-\vert\braket{\Phi_+}{\Phi_-}\vert} \leq \frac{1-\vert\braket{\psi_+}{\psi_-}\vert}{1-F(\rho_+,\rho_-)}.
\end{equation}
From the previous proposition we already know that the maximum
probability for $\ket{\psi_\pm} \rightarrow \ket{\Phi_\pm}$ is given
by Eq. (\ref{eqn-chefles}) and thus the upper bound holds
\begin{equation}
p \leq \min \left\lbrace \frac{1-\vert\braket{\psi_+}{\psi_-}\vert}{1-F(\rho_+,\rho_-)}, 1 \right\rbrace.
\end{equation}
This bound can be achieved by choosing the purifications $|\Phi_\pm\>$
which give the equality in equation (\ref{eqn-uhl}). The
transformation is balanced by the previous proposition.
\end{proof}
\end{proposition}

\section{Probability/fidelity tradeoff} \label{sec-probfid} 

Let us consider now the transformation
\begin{equation}
  \ket{\psi_\pm} \longrightarrow \ket{\varphi_\pm}, \quad \vert\braket{\varphi_+}{\varphi_-}\vert \leq \vert\braket{\psi_+}{\psi_-}\vert.
\end{equation}
By proposition \ref{thm-pairtrans} we know that it can be realized
\emph{exactly} only probabilistically. But if we allow also
\emph{approximate} transformations, realized by quantum operations
which transform $\ket{\psi_\pm}$ into states $\rho_\pm$ \emph{close}
to $\ket{\varphi_\pm}\bra{\varphi_\pm}$
\begin{equation}
\begin{split}
\ket{\psi_\pm}&\longrightarrow \rho_\pm = \mathcal{E}(\ketbra{\psi_\pm}{\psi_\pm})/p_\pm, \\
p_\pm &= \Tr(\mathcal{E}(\ketbra{\psi_\pm}{\psi_\pm})), 
\end{split}
\end{equation}
we may be able to implement the
transformation with greater probability, or even deterministically.

In general, there are two figures of merit characterizing the
transformation: (1) the probability of success, (2) the fidelity
between the target states and the states actually
obtained. Intuitively, the more we try to tilt the pair
$\ket{\psi_\pm}$ towards the target states, the less the
transformation is likely to happen.

The figures of merit are defined as follows
\begin{align}
p &= \min\left\{p_+, p_-\right\}\label{p}\\
F &= \min\left\{F(\ket{\varphi_+}\bra{\varphi_+},\rho_+),F(\ket{\varphi_-}\bra{\varphi_-},\rho_-)\right\}.\label{fidel}
\end{align}
where $p$ is the minimum probability and $F$ is the minimum fidelity
over the two states (a \emph{worst-case} criterion).  Each
transformation is characterized by a pair $(p,F)$, the set of all
transformations thus being in correspondence with a subset of $[0,1]
\times [0,1]$. Our task is to determine the frontier of this permitted
subset, thus finding the transformations maximizing both figures of
merit.

We can restrict our attention to approximated target states $\rho_\pm$
having the same two-dimensional support, equal to the linear span of
the target states $\ket{\varphi_\pm}$. In fact, exploiting the Kraus
representation for $\mathcal{E}$ \cite{kr83} (with Kraus operators
$K_j$) and defining the unnormalized states
$\ket{\beta_\pm^{(j)}}:=K_j\ket{\psi_\pm}$, we have
\begin{equation}
\rho_\pm = \frac1{p_\pm}{\sum_j \ket{\beta_\pm^{(j)}}\bra{\beta_\pm^{(j)}}}.
\end{equation}
We note that we can apply unitary operators $U_j$ after the Kraus
operators $K_j$ without altering the probabilities, obtaining new
states $\rho_\pm'$
\begin{equation}
\rho_\pm' = \frac1{p_\pm}{\sum_j U_j\ket{\beta_\pm^{(j)}} \bra{\beta_\pm^{(j)}} U_j^\dag}
\end{equation}
whose fidelity with the target states is
\begin{equation}
F(\ket{\varphi_\pm}\bra{\varphi_\pm}, \rho_\pm')
= \sqrt{\frac1{p_\pm} \sum_j |\bra{\varphi_\pm} U_j \ket{\beta_\pm^{(j)}}|^2}
\end{equation}
Thus, in order to have $F(\ket{\varphi_\pm}\bra{\varphi_\pm},
\rho_\pm') \geq F(\ket{\varphi_\pm}\bra{\varphi_\pm}, \rho_\pm)$ for
$\rho'_\pm$ supported on the span of $|\varphi_\pm\>$, we only need to
show that, given a pair of vectors $\ket{\beta_\pm}$, there is always
a unitary transformation $U$ moving $\ket{\beta_\pm}$ in the span of
$\ket{\varphi_\pm}$ such that
\begin{equation}
|\bra{\varphi_\pm}U\ket{\beta_\pm}| \geq |\braket{\varphi_\pm}{\beta_\pm}|.
\end{equation}
The operator $U$ can be costructed in the following way.

Let us consider the component of $\ket{\beta_+}$ orthogonal to
$\operatorname{Span}\{ \ket{\varphi_+}, \ket{\beta_-} \}$.  We rotate
it into the one-dimensional subspace of $\operatorname{Span}\{
\ket{\varphi_+},\ket{\varphi_-},\ket{\beta_-} \}$ orthogonal to
$\operatorname{Span}\{ \ket{\varphi_+}, \ket{\beta_-} \}$.  In this
way, we have moved the four vectors in a three-dimensional space
without changing the relevant scalar products
$|\<\varphi_\pm|\beta_\pm\>|$.  The intersection
$V=\operatorname{Span}\{ \ket{\varphi_+}, \ket{\varphi_-}
\}\cap\operatorname{Span}\{ \ket{\beta_+}, \ket{\beta_-} \}$ is
one-dimensional, thus we can rotate the components of
$\ket{\beta_\pm}$ orthogonal to $V$ into the one-dimensional subspace
of $\operatorname{Span}\{ \ket{\varphi_+}, \ket{\varphi_-} \}$
orthogonal to $V$.  This rotation leaves all vectors in
a two-dimensional space and increases the modulus of the scalar
products $|\braket{\varphi_\pm}{\beta_\pm}|$.

\begin{figure}
\begin{center}
  \includegraphics[width=\columnwidth]{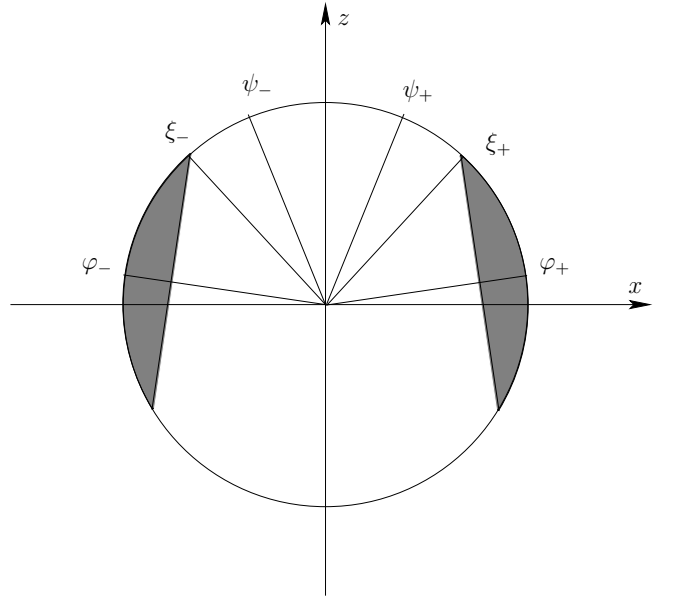}
\caption{\label{fig-circle}\small\linespread{1}\selectfont The section
  of the Bloch ball containing the initial pair $\ket{\psi_\pm}$ and
  the target pair $\ket{\varphi_\pm}$. The shadowed area contains all
  the states $\rho_\pm$ with fidelity $F(\rho_\pm, \ket{\varphi_\pm})
  \geq \vert\braket{\xi_\pm}{\varphi_\pm}\vert$.}
\end{center}
\end{figure}

\begin{figure}[h]
\includegraphics[width=\columnwidth]{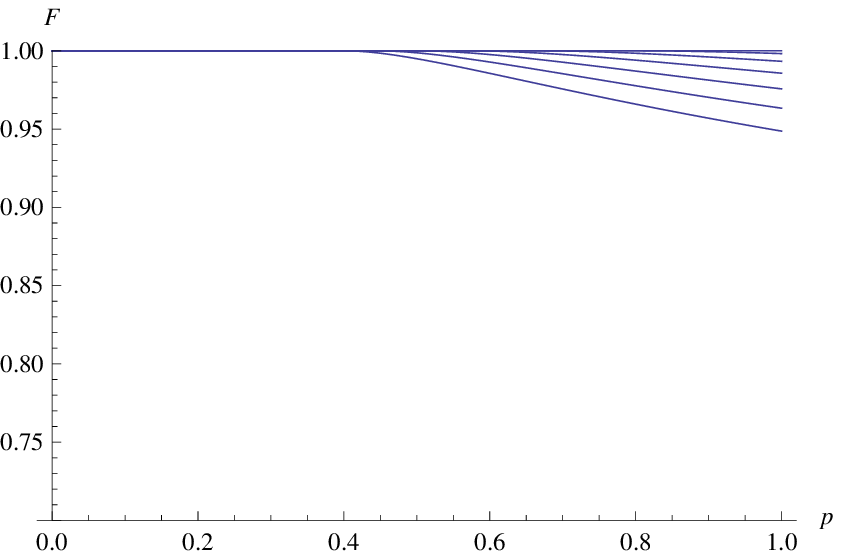}
\caption{\label{fig-pf1}\small\linespread{1}\selectfont Tradeoff
  curves $F(p)$ for $\vert\braket{\psi_+}{\psi_-}\vert=0.6$ and for
  $\vert\braket{\varphi_+}{\varphi_-}\vert$ ranging from 0 to 0.6, at
  intervals of $0.1$}
\end{figure}
\begin{figure}[h]
\includegraphics[width=\columnwidth]{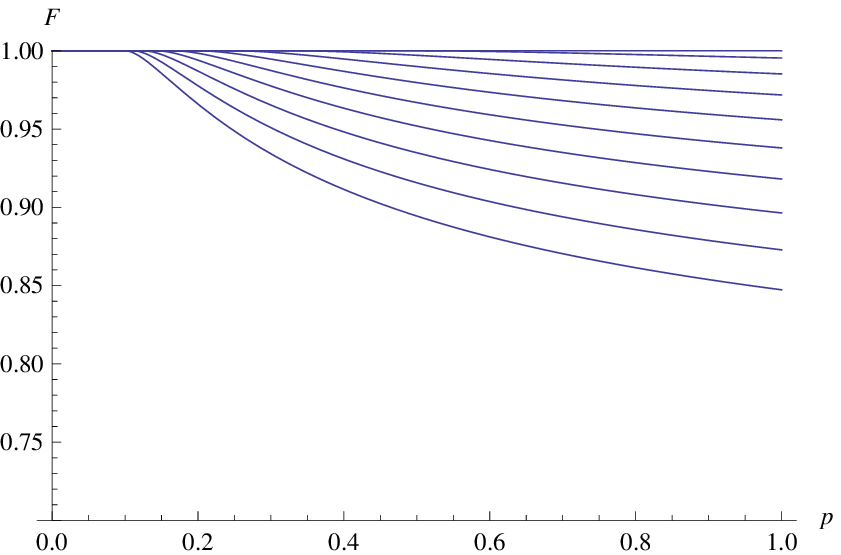}
\caption{\label{fig-pf2}\small\linespread{1}\selectfont Tradeoff
  curves $F(p)$ for $\vert\braket{\psi_+}{\psi_-}\vert=.9$ and for
  $\vert\braket{\varphi_+}{\varphi_-}\vert$ ranging from 0 to 0.9, at
  intervals of $0.1$}
\end{figure}
\begin{figure}[h]
\includegraphics[width=\columnwidth]{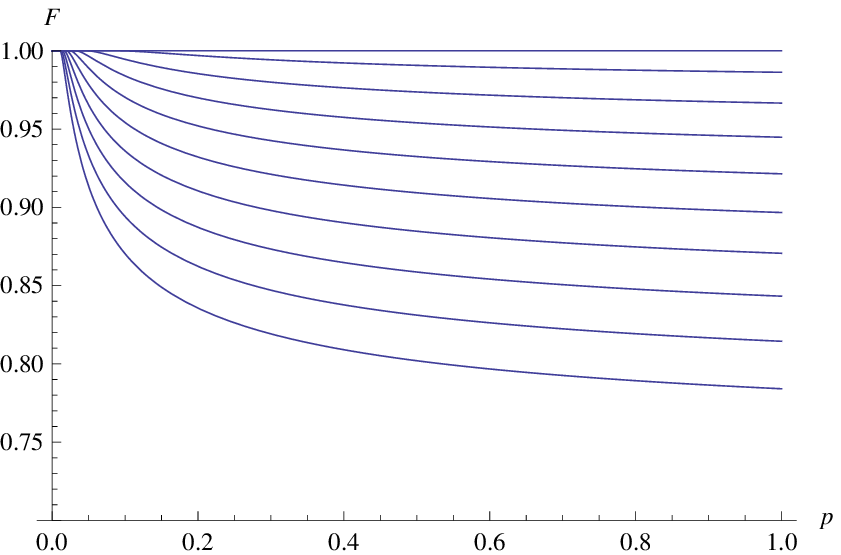}
\caption{\label{fig-pf3}\small\linespread{1}\selectfont Tradeoff
  curves $F(p)$ for $\vert\braket{\psi_+}{\psi_-}\vert=.99$ and for
  $\vert\braket{\varphi_+}{\varphi_-}\vert$ ranging from 0.09 to 0.99,
  at intervals of $0.1$}
\end{figure}

In the following we will then restrict to the span of
$|\varphi_\pm\>$, and it is convenient to use the Bloch representation
of states of bidimensional systems
\begin{equation}
\rho =   (I+\boldsymbol{r} \cdot \boldsymbol{\sigma})/2,
\end{equation}
where the \emph{Bloch vector} $\boldsymbol{r}=(x, y,
z)\in\mathbb{R}^3$ denotes a point in the unit ball
$|\boldsymbol{r}|\leq 1$ and $\boldsymbol{\sigma}=(\sigma_x, \sigma_y,
\sigma_z)$ is the vector of Pauli matrices. 
In the Bloch representation the fidelity between the states $\rho$ and
$\sigma$ (with Bloch vectors $\boldsymbol{r}_\rho$ and
$\boldsymbol{r}_\sigma$) becomes \cite{hu92}
\begin{equation}\label{eqn-hubner}
  F(\rho, \sigma) = \frac{\left(1+\boldsymbol{r}_\rho\cdot \boldsymbol{r}_\sigma + \sqrt{(1-\vert \boldsymbol{r}_\rho\vert^2)(1-\vert \boldsymbol{r}_\sigma\vert^2)}\right)^{\frac{1}{2}}}{\sqrt{2}},
\end{equation}
which, when one of the two states is pure simplifies as follows
\begin{equation}
F(\rho, \sigma) =\sqrt{\frac{1+\boldsymbol{r}_\rho\cdot \boldsymbol{r}_\sigma}{2}}.
\end{equation}

The angle between vectors $\boldsymbol{r}_{\rho_+}$
$\boldsymbol{r}_{\ket{\varphi_+}}$ and the angle between vectors
$\boldsymbol{r}_{\rho_-}$ $\boldsymbol{r}_{\ket{\varphi_-}}$ are both
minimized for the pair $\boldsymbol{r}_{\rho_\pm}$ coplanar with the
pair $\boldsymbol{r}_{\ket{\varphi_\pm}}$, and with the same symmetry
axis. This relative position of the couples of vectors can be achieved
by a rotation of the couple $\boldsymbol{r}_{\rho_\pm}$ in the Bloch
sphere, corresponding to a unitary transformation which leaves the
probabilities $p_\pm$ invariant. Now, for each operation $\mathcal{E}$
realizing a certain transformation
\begin{equation}
\mathcal{E}(\ketbra{\psi_{\pm}}{\psi_{\pm}})= p_{\pm}\rho_{\pm},
\end{equation}
where $\rho_{\pm}$ are coplanar with $\ket{\varphi_\pm}$, we can construct an operation $\mathcal{E}'$ acting in the following way
\begin{equation}
\mathcal{E}'(\ketbra{\psi_{\pm}}{\psi_{\pm}}) = \frac{1}{2}(p_\pm\rho_\pm+p_\mp \sigma_z \rho_\mp\sigma_z),
\end{equation}
where we have chosen the basis of the representation such that
$\sigma_z$ is the $\pi$-rotation around the symmetry axis of the pair
$\vert\varphi_\pm\rangle$, i.~e. $\sigma_z\vert\varphi_-\rangle
=\vert\varphi_+\rangle$. The second term in r.h.s. is simply the
``mirror image'' of $\mathcal{E}$.  This new quantum operation is
symmetric since $\sigma_z \mathcal{E}'
(\ketbra{\psi_{\pm}}{\psi_{\pm}}) \sigma_z = \mathcal{E}'
(\ketbra{\psi_{\mp}}{\psi_{\mp}}) = \mathcal{E}'
(\sigma_z\ketbra{\psi_{\pm}}{\psi_{\pm}}\sigma_z)$ and behaves better
than the original one w.r.t both figures of merit in Eqs. (\ref{p}) and
(\ref{fidel}), since
\begin{equation}
\begin{split}
&\Tr(\mathcal{E}'(\ketbra{\psi_{\pm}}{\psi_{\pm}}))=\frac{1}{2}(p_++p_-) \geq \min\lbrace
p_+,p_-\rbrace\\ 
&F(\ket{\varphi_\pm}, \mathcal{E}'(\ketbra{\psi_{\pm}}{\psi_{\pm}}))\geq\\
&\min\lbrace F(\ket{\varphi_+},\rho_+), F(\ket{\varphi_-},\rho_-)\rbrace.
\end{split}
\end{equation}

Thus, the frontier of the set of permitted couples $(p,F)$ can be
determined considering only symmetric transformations. Notice that
w.l.o.g.  we can assume the initial states $\ket{\psi_\pm}$ to be in
the symmetric configuration coplanar with $\ket{\varphi_\pm}$ (with
$\ket{\psi_+}$ close to $\ket{\varphi_+}$ and $\ket{\psi_-}$ close to
$\ket{\varphi_-}$), since this can be always achieved by a rotation of
the pair, corresponding to an additional unitary transformation, which
doesn't change probabilities.  In the $(p,F)$ plane this configuration
corresponds to the point $(1, f_0)$, where $f_0
:=\vert\braket{\psi_+}{\varphi_+}\vert \equiv
\vert\braket{\psi_-}{\varphi_-}\vert$.

Now, let $f$ be the fidelity we want to achieve, with $f_0 \leq f \leq
1$. Then, the set of possible final states $\rho_\pm$ compatible with
the constraint $F(\rho_\pm, \ket{\varphi_\pm})\geq f$ is the shadowed
area depicted in Fig. \ref{fig-circle}, where $\ket{\xi_\pm}$ are pure
states such that $\vert\braket{\xi_\pm}{\varphi_\pm}\vert:=f$.
We claim that among such states, the most probably attainable final
configuration is the pair $\ket{\xi_\pm}$. In order to prove the
claim, we need to prove that the probability
\begin{equation}\label{p2}
p = \frac{1-\vert\braket{\psi_+}{\psi_-}\vert}{1-F(\rho_+, \rho_-)}
\end{equation}
reaches the maximum at the pair $\ket{\xi_\pm}$, over any symmetric
pair $\rho_\pm$ inside the area. The fidelity $F(\rho_+, \rho_-)$ for
states compatible with the constraints $F(\rho_\pm,
\ket{\varphi_\pm})\geq f$ is maximized by the pair $\ket{\xi_\pm}$.
Indeed, the fidelity $F(\rho_+, \rho_-)$ for states of the form
$\rho_\pm = \frac{1}{2}(I\pm\beta\sigma_x + \gamma\sigma_z)$ can be
obtained from Eq. (\ref{eqn-hubner})
\begin{equation}
  F(\rho_+, \rho_-) = \sqrt{1-\beta^2}.
\end{equation}
This clearly shows that the optimal states maximizing probability
(\ref{p2}) are those minimizing $\beta$. The pair $\ket{\xi_\pm}$
satisfies this request, whence it is the most probable.

The remaining part of the optimal tradeoff curve can now be completed
quite easily; we only need to sweep the pure states in the arc between
$\ket{\psi_\pm}$ and $\ket{\varphi_\pm}$ to obtain the points in the
$(p,F)$-plane connecting $(1, f_0)$ and $(p_0, 1)$, where
$p_0=(1-\vert\braket{\psi_+}{\psi_-}\vert) /
(1-\vert\braket{\varphi_+}{\varphi_-}\vert)$. After a little
trigonometry, we obtain the explicit expression for this part of the
curve
\begin{equation}
F(p)=\cos\left[\frac{\arccos\vert\braket{\varphi_+}{\varphi_-}\vert - \arccos\left(1-\frac{1-\vert\braket{\psi_+}{\psi_-}\vert}{p} \right)}{2}\right].
\end{equation}

In figures \ref{fig-pf1},\ref{fig-pf2},\ref{fig-pf3} we plot these
curves for different values of the fidelities
$|\braket{\psi_+}{\psi_-}|$ and $|\braket{\varphi_+}{\varphi_-}|$.


\section{Tradeoff for the inversion of a quantum operation}

Suppose we want to know whether a given quantum operation $\mathcal{E}$
can be inverted deterministically on some subspace
$\mathcal{L}\subseteq\mathcal{H}$, in other words whether there is a quantum
channel $\mathcal{R}$ such that
\begin{equation}
\rho \longrightarrow \rho' = \frac{\mathcal{E}(\rho)}{\Tr(\mathcal{E}(\rho))} \longrightarrow \mathcal{R}(\rho') = \rho
\end{equation}
for every $\rho$ with $\supp(\rho) \subseteq \mathcal{L}$. Necessary
and sufficient conditions for this inversion have been proved in
Ref. \cite{kl96}, while in Ref. \cite{sn96} an equivalent condition
based on information-theoretical quantities such as entropy and
coherent information is provided. If the quantum operation cannot be
inverted by a channel, or the inversion is not required to be perfect,
it is still possible to achieve an approximate inversion which brings
$\rho'$ \emph{close} to $\rho$. Such ``closeness'' has been quantified
in Ref. \cite{sw02}, whenever $\mathcal{E}$ is a channel, and in
Ref. \cite{bhh07} for general quantum operations.

In the present paper we explore the possibility of
\emph{probabilistic} inversions, including \emph{exact} inversions as
a particular case. In the following we will focus on a two-dimensional
system undergoing an atomic quantum operation
\begin{equation}
\mathcal{E}(\rho)=M\rho M^\dag,
\end{equation}
where $M$ is a \emph{contraction}, i.e. satisfying $|\!|M|\!|\leq
1$. Using the polar decomposition $M=UP$ with unitary $U$ and $P\geq
0$, w.~l.~o.~g. we can take $M=M_\beta\geq 0$ with the following
matrix representation
\begin{equation}
M_\beta=\left(
\begin{array}{cc}
1 & 0 \\ 
0 & \beta
\end{array}
\right)
\end{equation}
where $\beta$, $0\leq\beta\leq 1$, is the smallest singular value. The
largest singular value can be fixed at $1$ up to an overall
probability rescaling independent of the state (we assume that the
quantum operation has happened).

We will consider two case studies with a given set $\mathcal{D}$ of
initial states, and a given set $\mathcal{Q}$ of quantum operations
inverting $M_\beta$ approximately. After the transformation on the
state $\rho\in\mathcal{D}$
\begin{equation}
\rho'=\frac{M_\beta\rho M_\beta}{\Tr(\rho M_\beta^2)},
\end{equation}
a following ``inverting'' quantum operation
$\mathcal{R}\in\mathcal{Q}$ leaves the system in the state
\begin{equation}
\rho''=\frac{\mathcal{R}(\rho')}{\Tr(\mathcal{R}(\rho'))}.
\end{equation}
The quality of the inversion is assessed by two figures of merit: i)
the probability of success
\begin{equation}
p(\mathcal{R};\rho)=\Tr(\mathcal{R}(\rho'));
\end{equation}
ii) the fidelity with the initial state
\begin{equation}
f(\mathcal{R};\rho)=F(\rho,\rho'').
\end{equation}
In order to keep the probability of success above some threshold
$\overline{p}$ we consider only the subset
$\mathcal{Q}_{\overline{p}}\subseteq\mathcal{Q}$ whose elements
$\mathcal{R}$ satisfy the constraint:
\begin{equation}
p(\mathcal{R};\rho)\geq \overline{p}, \quad \forall\rho\in\mathcal{D}.
\end{equation}
In a worst-case criterion we have to choose the inversion
$\overline{\mathcal{R}}\in\mathcal{Q}_{\overline{p}}$ maximizing the
minimum fidelity over $\mathcal{D}$
\begin{equation}
  \overline{\mathcal{R}} = \arg\max_{\mathcal{R}\in\mathcal{Q}_{\overline{p}}} \min_{\rho\in\mathcal{D}} f(\mathcal{R};\rho).
\end{equation}
This gives the point $(\overline{p}, \overline{F})$, with
$\overline{F}=\min_{\rho\in\mathcal{D}}
f(\overline{\mathcal{R}};\rho)$, in the $(p,F)$ plane. The tradeoff
curve is obtained varying $\overline{p}$ in the interval $[0,1]$. In
this way we obtain a curve $F=F(p)$ giving the minimum fidelity over
$\mathcal{D}$ achievable with probability of success at least $p$.

\subsection{Semiclassical case}

The set of input states $\mathcal{D}=\{\rho_x \}$ consists of all
density operators jointly diagonal with the contraction
\begin{equation}
\rho_x =\left(
\begin{array}{cc}
x & 0 \\ 
0 & 1-x
\end{array}
\right), \quad 0\leq x\leq 1,
\end{equation}
while the set of possible inversions $\mathcal{Q}=\{N_\gamma\}$ consists of the diagonal contractions
\begin{equation}
N_\gamma =\left(
\begin{array}{cc}
\gamma & 0 \\ 
0 & 1
\end{array}
\right), \quad \beta\leq\gamma\leq 1.
\end{equation}
The unit-fidelity case is the matrix inverse (rescaled in order to
keep it a contraction) $N_\beta = M_\beta^{-1}/\Vert M_\beta^{-1}
\Vert$ \cite{ noteflip}.


\begin{figure}[t]
\includegraphics[width=\columnwidth]{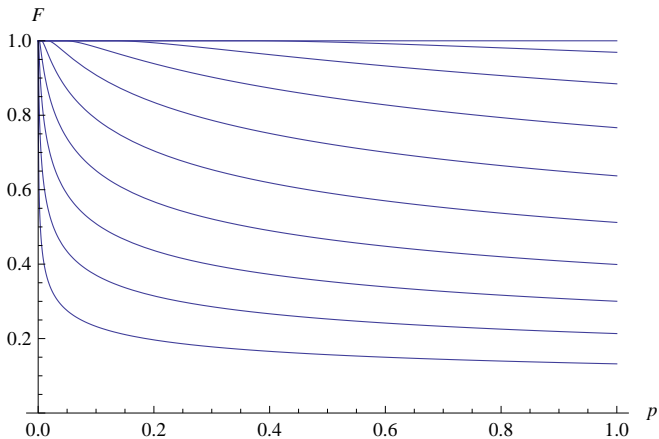}
\caption{\label{fig-invtrade}\small\linespread{1}\selectfont Tradeoff
  curves for the semiclassical case, far various $\beta$. Each curve
  gives the minimum guaranteed fidelity in the inversion of $M_\beta$,
  as a function of the minimum probability $p$ of success over all the
  initial states. Curves from the bottom to the top correspond to
  increasing values of $\beta$, ranging from 0.1 to 1.}
\end{figure}

The states $\rho_x'$ and $\rho_x''$ are easily computed
\begin{align}
\rho_x'=\frac{1}{x+\beta^2(1-x)}
\left(\begin{array}{cc}
x & 0\\
0 & \beta^2(1-x)
\end{array}\right),\\
\rho_x''=\frac{1}{\gamma^2x+\beta^2(1-x)}
\left(\begin{array}{cc}
\gamma^2 x & 0\\
0 & \beta^2(1-x)
\end{array}\right),
\end{align}
and so are the probability and the fidelity
\begin{align}
p(N_\gamma;\rho_x)=\frac{\gamma^2x+\beta^2(1-x)}{x+\beta^2(1-x)},\\
f(N_\gamma;\rho_x)=\frac{\gamma x+\beta(1-x)}{\sqrt{\gamma^2x+\beta^2(1-x)}}.
\end{align}
By inspection of these expressions one can see that the set $\mathcal{Q}_{\overline{p}}$ is
\begin{equation}
  \mathcal{Q}_{\overline{p}}=\left\{N_\gamma,\quad \gamma^2\geq\overline{p}\right\}
\end{equation}
and that
\begin{equation}
\arg\max_{N_\gamma \in\mathcal{Q}_{\overline{p}}} \min_{\rho_x} f(N_\gamma;\rho_x)=N_{\sqrt{\overline{p}}}
\end{equation}

The corresponding tradeoff curves are plotted in Fig.
\ref{fig-invtrade} for various $\beta$. The uppermost curves are
obtained when $\beta$ approaches $1$, i.e. when $M_\beta$ is near to
the identity (clearly, in this case there is almost no need of
inversion). On the other hand, as $\beta$ goes to zero $M_\beta$
approaches an orthogonal projector which, in our worst-case criterion,
cannot be inverted with nonvanishing minimum fidelity.

\subsection{Quantum case}
We consider a set of two non-orthogonal states
$\mathcal{D}=\lbrace\ket{\psi_\pm}\rbrace$, and we let $\mathcal{Q}$
to be the set of \emph{all} quantum operations. The states after the
first transformation are
\begin{equation}
  \ket{\psi_\pm'} = \frac{M_\beta\ket{\psi_\pm}}{\vert\!\vert M_\beta\ket{\psi_\pm}\vert\!\vert}.
\end{equation}
The required inversion is then
\begin{equation}
\ket{\psi_\pm'}\longrightarrow \ket{\psi_\pm}
\end{equation}
which we already studied in Section \ref{sec-probfid}.

\medskip
\section{Conclusions}
After generalizing the state-transformation probability formula of
Ref. \cite{cb98} to mixed target states, we derived a
probability-fidelity tradeoff for a varying quantum operation with
fixed input-output states. We have then presented the first tradeoff
between the probability and the fidelity in the inversion of a quantum
operation in a semiclassical and in a quantum case.

\acknowledgments This work has been supported by the EC through the
project CORNER.


\begin{thebibliography}{99}

\bibitem{fp96}
C. A. Fuchs and A. Peres, 
Phys. Rev. A \textbf{53}, 2038 (1996).
\bibitem{fu98}
C. A. Fuchs,
Fortschr. Phys. \textbf{46}, 535 (1998).
\bibitem{ba00}
H. Barnum, Report University of Bristol (2000), quant-ph/0205155. 
\bibitem{banaszek01.prl} K.~Banaszek, Phys. Rev. Lett. {\bf 86}, 1366 (2001).
\bibitem{banaszek01.pra} K. Banaszek and I. Devetak, Phys. Rev. A
  {\bf 64}, 052307 (2001).
\bibitem{mista05.pra} L. Mi{\v s}ta Jr., J. Fiur\'a{\v s}ek, and R.
  Filip, Phys.  Rev. A {\bf 72}, 012311 (2005).
\bibitem{max}M. F. Sacchi, Phys. Rev. Lett. {\bf 96}, 220502 (2006). 
\bibitem{cv}U. L. Andersen, M. Sabuncu, R. Filip, and G. Leuchs, 
Phys. Rev. Lett. {\bf 96}, 020409 (2006).
\bibitem{paris}M. G. Genoni and M. G. A. Paris, Phys. Rev. A {\bf 74}, 012301 (2006).  
\bibitem{bs06}
F. Buscemi and M. F. Sacchi, 
Phys. Rev. A \textbf{74}, 052320 (2006).
\bibitem{maxun}M. F. Sacchi, Phys. Rev. A {\bf 75}, 012306 (2007). 

\bibitem{da03}
G. M. D'Ariano, 
Fortschr. Phys. \textbf{51}, 318 (2003).


\bibitem{cb98}
A. Chefles and S. M. Barnett,
J. Phys. A \textbf{31}, 10097 (1998).

\bibitem{uh76}
A. Uhlmann, 
Rep. Mat. Phys. \textbf{9}, 273 (1976).

\bibitem{ozawa} 
M.~Ozawa, J. Math. Phys. {\bf 25}, 79 (1984).


\bibitem{kr83}
K. Kraus,
\emph{States, Effects and Operations. Fundamental Notions of Quantum
  Theory} (Springer-Verlag, Berlin Heidelberg, 1983).



\bibitem{hu92}
M. H\"ubner, 
Phys. Lett. A \textbf{163}, 239 (1992).

\bibitem{kl96}
  E. Knill and R. Laflamme,
\textsf{arXiv:quant-ph/9604034} (1996).

\bibitem{sn96}
B. Schumacher and M. Nielsen,
Phys. Rev. A \textbf{54}, 2629 (1996).


\bibitem{sw02}
B. Schumacher and M. Westmoreland, 
Quantum Information Processing \textbf{1}, 5 (2002).


\bibitem{bhh07}
F. Buscemi, M. Hayashi and M. Horodecki, 
Phys. Rev. Lett. \textbf{100}, 210504 (2008). 


\bibitem{noteflip} The ``flipping'' contractions $P_\delta =\left(\begin{array}{cc}
0 & \delta \\ 
1 & 0
\end{array}
\right), \quad 0\leq\delta\leq 1$ mantain the diagonal form of the states. However, they do not
improve both the fidelity and the probability of inversion. 














\end{thebibliography}
\end{document}